\documentclass{aa}

\usepackage{epsfig}

\newbox\grsign \setbox\grsign=\hbox{$>$} \newdimen\grdimen \grdimen=\ht\grsign
\newbox\simlessbox \newbox\simgreatbox \newbox\simpropbox \newbox\wtildebox 
\setbox\simgreatbox=\hbox{\raise.5ex\hbox{$>$}\llap
     {\lower.5ex\hbox{$\sim$}}}\ht1=\grdimen\dp1=0pt
\setbox\simlessbox=\hbox{\raise.5ex\hbox{$<$}\llap
     {\lower.5ex\hbox{$\sim$}}}\ht2=\grdimen\dp2=0pt

\newcommand{\Msun}{\mbox{$M_{\odot}$}}
\newcommand{\msun}{\mbox{$M_{\odot}$}}
\newcommand{\Msunpc}{\mbox{$M_{\odot}$~pc$^{-2}$}}
\newcommand{\Mcl}{\mbox{$M_{\rm cl}$}}

\newcommand{\be}{\mbox{\begin{equation}}}
\newcommand{\ee}{\mbox{\end{equation}}}

\newcommand{\tdis}{\mbox{$t_{\rm dis}$}}
\newcommand{\trxt}{\mbox{$t_{\rm rt}$}}

\newcommand{\tcross}{\mbox{$t_{\rm cr}$}}

\newcommand{\rhoamb}{\mbox{$\rho_{\rm amb}$}}
\newcommand{\mmean}{\mbox{$\langle m \rangle$}}

\newcommand{\kms}{\mbox{km~s$^{-1}$}}
\newcommand{\rh}{\mbox{$r_{\rm hm}$}}

\begin{document}

\title{
Disruption time scales of star clusters in different galaxies}

\author{Henny J.G.L.M. Lamers \inst{1,2}, Mark Gieles\inst{1} 
        and Simon F. Portegies Zwart \inst{3,4} 
        }

\institute { 
                 {Astronomical Institute, Utrecht University, 
                 Princetonplein 5, NL-3584CC Utrecht, the Netherlands
                 {\tt  lamers@astro.uu.nl,  gieles@astro.uu.nl}}
            \and  {SRON Laboratory for Space Research, Sorbonnelaan 2,
                 NL-3584CC, Utrecht, the Netherlands}
            \and {Astronomical Institute, University of Amsterdam,
                 Kruislaan 403, NL-1098SJ, Amsterdam, the Netherlands
                 {\tt spz@science.uva.nl}}
            \and {Informatics Institute, University of Amsterdam,
                 Kruislaan 403, NL-1098SJ, Amsterdam, the Netherlands
                 }
            }

\date{Received date ; accepted date}

\offprints{H. J. G. L. M. Lamers}

\abstract{
The observed average lifetime of the population of star clusters in the
Solar  Neighbourhood, the Small Magellanic Cloud and in selected
regions of M51 and M33 is compared with simple theoretical predictions and with
the results of $N$-body simulations.
The empirically derived lifetimes (or disruption times) 
of star clusters depend on their initial mass
as $\tdis^{\rm emp} \propto \Mcl^{0.60}$ in all four galaxies.
$N$-body simulations have shown that the predicted
disruption time of clusters in a tidal field scales as 
$\tdis^{\rm pred} \propto t_{\rm rh}^{0.75} t_{\rm cr}^{0.25}$, where 
$t_{\rm rh}$ is the initial half-mass relaxation time and $t_{\rm cr}$
is the crossing time for a cluster in equilibrium. We show that
this can be approximated accurately by 
$\tdis^{\rm pred} \propto M_{\rm cl}^{0.62}$ for
clusters in the mass range of about $10^3$ to $10^6$ \Msun, in
excellent agreement with the observations.
Observations of clusters in different extragalactic environments show
that $\tdis$ also depends on the ambient density in the galaxies where 
the clusters reside. Linear analysis predicts that 
the disruption time will depend on the ambient density of the cluster
environment as $\tdis \propto \rho_{\rm amb}^{-1/2}$. 
This relation is consistent with $N$-body simulations. The
empirically derived
disruption times of clusters in the Solar Neighbourhood, in the SMC
and in M33 agree with these predictions. The best fitting expression
for the disruption time is \\
~~~~~~~~~~~~~~~~~~~~~ $\tdis=C_{\rm env} (\Mcl/10^4 \Msun)^{0.62} 
(\rhoamb / \Msun {\rm pc}^{-3})^{-0.5}$\\
where $\Mcl$ is the initial mass of the cluster 
and $C_{\rm env} \simeq 300 - 800$ Myr. The disruption times of star 
clusters in M51 within 1 - 5 kpc from the nucleus, is shorter than predicted 
by about an order of magnitude. This
discrepancy might be due to the strong tidal field variations in M51, 
caused by the strong density contrast between the spiral arms and 
interarm regions, or to the disruptive forces from giant molecular clouds.

\keywords{
Galaxy: open clusters --
Galaxy: solar neighborhood --
Galaxies: individual: M33 --
Galaxies: individual: M51 --
Galaxies: individual: SMC --
Galaxies: star clusters --
}
}

\authorrunning{H.J.G.L.M. Lamers et al.}
\titlerunning{Disruption of star clusters in different galaxies} 

\maketitle


\section{Introduction}
\label{sect:1}

The age distribution of star clusters in the disk of the Milky Way, can only be
explained if galactic clusters disrupt on a time scale on the order
of a few times $10^8$ years (Oort 1957; Wielen 1971, 1988).  Clusters
in the LMC and SMC survive longer than those in the solar
neighborhood (e.g.  Elson \& Fall 1985,1988; Hodge 1987).

 The disruption time of clusters is expected to depend on both the
{\it internal cluster conditions}, such as the initial mass, density
and velocity dispersion and the stellar initial mass function (IMF),
and on the {\it external conditions}, such as the orbit in the galaxy
and tidal heating by encounters with for example giant molecular clouds.
 Recently, Boutloukos
\& Lamers (2003, hereafter called BL03) have derived an empirical
expression for the disruption time of clusters as a function of their
initial mass in selected regions in four galaxies. They found large differences
in the disruption times between these regions.  This allows a crucial
test of the theoretical disruption times of clusters predicted with
$N$-body simulations.

The purpose of this paper is three-fold:\\
(a) to explain the empirical dependence of the cluster disruption times on 
cluster mass; \\
(b) to explain the strong dependence, found by BL03 of the disruption time
on the conditions in the host galaxies; \\
(c) to confront the predicted disruption theory with 
the empirically derived  disruption relations.

In Sect. 2 we describe the predicted dependence of the disruption
times of star clusters on their initial mass and on their environment.
In Sect. 3 we review the method used for the determination
of the disruption times based on a statistical analysis of 
large cluster samples and the resulting expressions
for the disruption times in four galaxies.
In Sect. 4 we discuss the observed dependence of the disruption times
on the environment of the clusters in their host galaxies and we
compare it with the predictions. We discuss the peculiar case of 
the clusters in M51. The conclusions are in Sect 5.


\section{The predicted dependence of the disruption time on  cluster
parameters and environment}

The evolution of an isolated star cluster is characterized by three distinct
phases: 
(1) infant mortality (the dissolution of unbound clusters) within
about 10 Myr (Bastian et al. 2004; Whitmore 2004),
(2) stellar evolution dominates the cluster mass loss in the
first $\sim 10^8$ years (see Takahashi \& Portegies Zwart
2000)\nocite{2000ApJ...535..759T} followed by (3) a relaxation
dominated phase in which the cluster mass loss is driven by its
internal dynamical evolution and the external influence of the tidal
field of the host galaxy. If the cluster survives the first two phases, its
lifetime will be dominated by the third phase as it generally lasts
much longer, until complete dissolution. During this episode mass loss
is about constant with time and appears to depend weakly on the
density profile of the cluster 
(Spitzer 1987; Portegies Zwart et al.\, 2001b, but see
Portegies Zwart et al.\, 1998, for some complications near the end of
the lifetime of the star cluster).  During the third phase there are
two important effects which drive the cluster dissolution. These are
internal two-body relaxation and the interaction between the cluster
and the tidal field of the host Galaxy.

Internal two-body relaxation drives the cluster evaporation on a very
long time scale of many initial half-mass relaxation times (Baumgardt
et al. 2002)\nocite{2002MNRAS.336.1069B}. 
The half-mass relaxation time scale can be written as (Spitzer 1958):

\begin{equation}
t_{\rm rh} = 0.138 \frac{N^{1/2}r_{\rm hm}^{3/2}}{\mmean^{1/2}
   G^{1/2}\ln{\Lambda}}~ 
\propto ~\frac{N}{\ln \Lambda} ~ \left( \frac{\rh^{3}}{ G \Mcl} \right)^{1/2}
\label{eq:relaxation}
\end{equation}
where $N$ is the initial number of stars in the clusters, 
$\mmean= \Mcl/N$ is the mean mass of the stars and $\rh$ is the
half-mass radius. The Coulomb logarithm $\Lambda$ is approximately 
proportional to $N$ (see Gierz \& Heggie 1996; Spinnato et
al. 2003). 

The mass loss rate for clusters in a tidal field is much higher
and the lifetime shorter 
than for clusters in isolation. 
\cite{baumgardt01} found that for clusters in a tidal field
the disruption time depends not only on the relaxation time but also 
on the crossing time \tcross\ because, due to their circulation within
the cluster, stars with sufficient energy
still need a considerable time to reach the outskirts of the cluster
from where they can escape. He showed that the disruption time can be 
approximated as

\begin{equation}
t_{\rm dis} \simeq t_{\rm rh}^x\>t_{\rm cr}^{1-x}
\label{eq:tdispred}
\end{equation}
where the crossing time for a cluster in equilibrium is

\begin{equation}
t_{\rm cr} \propto r_{\rm hm}^{3/2}/\sqrt{G \Mcl}.
\label{eq:tcross}
\end{equation}
and so $t_{\rm cr} \propto t_{\rm rh} \ln\Lambda/N$.

Baumgardt (2001) showed on theoretical arguments that
$x$ is expected to be about 3/4. A more accurate estimate of the
value of $0<x<1$ can be found empirically from $N$-body simulations (see
below).
Combining Eqs. 
\ref{eq:relaxation}, \ref{eq:tdispred} and \ref{eq:tcross} 
an expression for the
disruption time can be derived of the form

\begin{equation}
t_{\rm dis} = C\left[\frac{N}{\ln{(\gamma N)}}\right]^x
\left[\frac{\rh^{3}}{G \Mcl}\right]^{1/2}
\label{eq:tdis2}
\end{equation}
where we replaced $\Lambda$ with $\gamma N$, with $0.01 <\gamma <0.4$.
 
Baumgardt \& Makino (2003, hereafter BM03) have performed 
an extensive set of simulations of
clusters in the Galactic halo tidal field and fitted the results to find
the constant $C$ and the value of $x$ in the approximation of
Eq. \ref{eq:tdis2}. They used a logarithmic potential of the form
$\phi(R_{\rm G}) = V_{\rm G}^2\ln(R_{\rm G})$, where $R_{\rm G}$ is
the distance
 to the galactic center and $V_{\rm G}$ is the circular velocity.
 Their clusters started with a tidal radius equal to the tidal radius
 of the external field,

\begin{equation}
r_{\rm t} = \left(\frac{G \Mcl}{2V_{\rm G}^2}\right)^{1/3}R_{\rm G}^{2/3}
\label{eq:rtidal}
\end{equation}
If we assume the tidal radius scales linearly with the half-mass
radius, which holds for clusters with the same density profile,
Eq. \ref{eq:rtidal} can be combined with Eq. \ref{eq:tdis2},
which yields

\begin{equation}
\frac{t_{\rm dis}}{\rm Myr} = \beta\left[\frac{N}{\ln{(\gamma N})}\right]^x\frac{R_{\rm G}}{\rm kpc}\left(\frac{V_{\rm G}}{220\,\kms}\right)^{-1}
\label{eq:tdis3}
\end{equation}
where $\beta$ is a constant whose value can be found empirically.

\begin{figure}[!h]
\includegraphics[width=8cm]{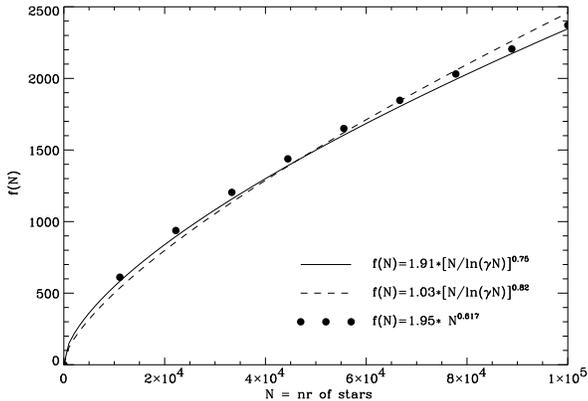}
\caption{The function $\beta\{N/\ln(0.02N)\}^x$ of
  equation~\ref{eq:tdis3} for the two combinations of
$\beta$ and $x$  for clusters with $W_0$ = 5 (solid line) and 7
(dashed) from BM03 and the power law approximation 
$1.95~N^{0.617}$ (dots).}
\label{fig:func}
\end{figure}

BM03  fitted this relation to the results of their
simulations of clusters at different distances from the Galactic
center.  For 
clusters that initially followed a King (1966) profile with
central concentration $W_0=5.0$ they found $\beta$ = 1.91 and $x$ = 0.75.
For clusters with $W_0=7.0$ they found $\beta$ = 1.03 and $x$ = 0.82.
For $\gamma$ a value of 0.02 was adopted. 
We found that the factor $\beta \{N/ \ln(0.02 N)\}^{x}$ in 
Eq. \ref{eq:tdis3}  
for both combinations of $\beta$ and $x$ can be approximated
by a power law $1.95 N^{0.62}$ with high accuracy.
This is illustrated in Fig.~\ref{fig:func}. 

The density of the clusters depends on the ambient density which
depends on the adopted potential field. The {\it ambient
  density}, for the
potential field adopted in the study by BM03
can be found by applying Poisson's law which gives

\begin{equation}
\rho_{\rm amb} = \frac{1}{4\pi G}\left(\frac{V_{\rm G}}{R_{\rm G}}\right)^2
\label{eq:poisson}
\end{equation}
Combining this with Eq. \ref{eq:tdis3}, and assuming an initial mean
stellar mass of $\mmean =
0.54 \Msun$, as done by BM03 based on the 
stellar IMF of Kroupa (2001), we can write the disruption time 
as a function of cluster mass and local ambient density:

\begin{equation}
t_{\rm dis} \simeq 810\>\left(\frac{M_{\rm
      cl}}{10^4\>\msun}\right)^{0.62}\>\rho_{\rm amb}^{-1/2}~~~~~{\rm Myr}
\label{eq:tdisBM}
\end{equation}
with $\Mcl$ in units of $\msun$ and $\rho_{\rm amb}
$ in  $\msun$pc$^{-3}$.
The constant indicates the time when 95 per cent of the initial
cluster mass is lost.
If the initial mean stellar mass is 0.70 \Msun,  predicted by the 
stellar IMF from Scalo (1986), the constant decreases from
810 to 690. 
(For clusters in elliptical orbits with ellipticity $\epsilon$ 
moving within a logarithmic potential field the
disruption time is shorter by factor $1-\epsilon$, 
if $\rho_{\rm amb}$ is the density at the apogalactic radius.) 
The positive exponent 0.62 indicates that the disruption time
increases with mass, as expected.

\begin{figure}
\centerline{\psfig{figure=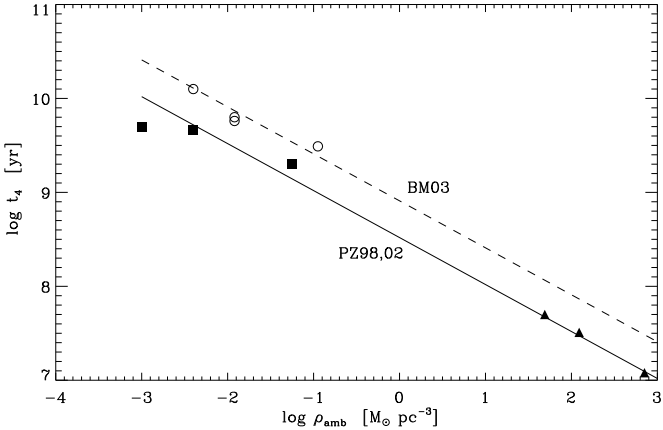,width=10.0cm,angle=0}}
\caption[]{The predicted dependence of the disruption time, $t_4$, of 
clusters of initial mass of $10^4$ \Msun\ on the mean ambient
density, \rhoamb.
The symbols refer to $N$-body simulations; filled squares
 are from Portegies Zwart et
al. (1998, from left to right models IR16, IC16 and
FH16)\nocite{1998A&A...337..363P}, filled triangles are calculations for star
clusters near the Galactic center from Portegies Zwart et al (2002,
from left to right models R150W4, R90W4 and
R34W4)\nocite{2002ApJ...565..265P}. The open circles 
represent models from
BM03 (from left to right models V, I, II and IV
with 16384 stars)\nocite{2003MNRAS.340..227B}.  
The dashed line is the relation predicted by BM03,
Eq. \ref{eq:tdisBM}, and the full line is the relation
predicted by Portegies Zwart et al. }
\label{fig:t4predicted}
\end{figure}

The results of $N$-body simulations and their predicted dependence on the
ambient density are plotted in Fig. \ref{fig:t4predicted}. 
These are based on  $N$-body simulations of
clusters of $10^4$ \Msun\ from different authors. 
These simulations are from Portegies Zwart et
al. (1998; 2002), \nocite{1998A&A...337..363P, 2002ApJ...565..265P}
and BM03 \nocite{2003MNRAS.340..227B}.  All
simulations were performed with {\tt NBODY4} 
(Aarseth 1999; BM03)\nocite{2003MNRAS.340..227B} and the {\tt Starlab}
(Portegies Zwart et al.\, 2001a)\nocite{2001MNRAS.321..199P}\footnote{see
also {\tt http://www.manybody.org}} software environment with both codes
running on the GRAPE family of special purpose computers (Makino
2003).\nocite{2003PASJ...55.1163M} 
The models  by Portegies Zwart et al. (2002) were calculated for star
clusters near the Galactic center, i.e. they are representative for 
a high density ambient medium.

The dashed line shown in this figure represents the mean relation
derived by BM03  and described by 
Eq. (\ref{eq:tdisBM}). This line passes through the data 
of their individual models. 

The full line is the mean relation
predicted by Portegies Zwart et al. (2001b). It is derived 
from the argument that the lifetime of a cluster will depend on the 
two-body relaxation time near the tidal radius as $\tdis  \simeq 0.3 \trxt$.
The tidal radius can be obtained from Eq. \ref{eq:rtidal}, but the
required average cluster density $\rho_{\rm cl}$ is not readily
available from the observations. Since the cluster is in equilibrium
with its environment one can estimate $\rho_{\rm cl}/\rho_{\rm amb} \simeq   
3$ for a standard Roche-solution 
by assuming that the clusters follow a King (1966) model and
that the tidal radius of the King model equals the Jacobi radius of
the star cluster in the tidal field of the Galaxy
(see e.g. Binney \& Tremaine 1994). 
This predicts that the constant in Eq. (\ref{eq:tdisBM})
is about 330 Myr. This is 0.4 times as small as
the value derived from the models by BM03 .
The line passes through the data of the individual models of
Portegies Zwart et al. (1998, 2002).

The simulations by Portegies Zwart et al. (1998, 2002) result in
shorter disruption times than the simulations by BM03, by about a
factor of 2.5. This difference can be explained by variations in the
initial conditions and between the two codes.  The initial mass
function used in Portegies Zwart et al.
(2002)\nocite{2002ApJ...565..265P} is taken from Scalo
(1986)\nocite{scalo86} and has a mean stellar mass of $\mmean \simeq
0.7$\msun. BM03 adopted the initial mass function of Kroupa (2001),
which has a mean stellar mass of 0.54 \Msun.  Portegies Zwart et
al. (1998) and BM03 use King (1966) models for the initial density
distribution, whereas Portegies Zwart et al (2002) adopt self
consistent Heggie \& Ramamani (1995) models.  But also the stellar
evolution in the calculations are slightly different: the stellar
evolution in BM03 is based on Pols et al. (1998) whereas
the calculations of Portegies Zwart et al. use the stellar evolution 
predicted by Eggleton et al. (1989).   
Moreover, the end of the
cluster lifetime is defined differently in the various calculations:
BM03 and Portegies Zwart et al. (2002) uses the time when 5 percent of
the initial mass remains, whereas Portegies Zwart et al. (1998) use
the time when 10 percent of the initial mass remains. The most
important difference, however, is in the treatments of the tidal field
and the way stars escape from the cluster.  Portegies Zwart et
al. (1998) did not include a full tidal potential, but a simple
cut-off radius. This easily leads to a shorter disruption time for the
cluster. BM03 and Portegies Zwart et al (2002), on the other hand
adopt a self-consistent tidal field up to the quadrupole moment.\footnote{
In an extensive {\em collaborative} experiment Heggie et
al. (1998) perform a detailed comparison between the N-body codes  
used by Portegies Zwart and Baumgardt (and others). The difference
in cluster lifetime is between 20 and 40\%. (see 
{\tt http://www.maths.ed.ac.uk/$\sim$douglas/experiment.html})}

A few notes about these estimate:\\
(a) The constant of expression (\ref{eq:tdisBM}) was derived for 
clusters moving 
in circular orbits in the Galactic potential field.  
Clusters in elliptical or inclined orbits will experience disk or 
bulge shocking that 
reduces their lifetime (Spitzer 1987, Zhang \& Fall 1999; BM03).
However the proportionality with mass and density remains the same.\\
(b) Encounters with giant molecular clouds (GMCs) or spiral density
waves are not included in these calculations. In situations where
these effects are likely to occur, e.g. for clusters in orbits in a 
galactic plane, the constant will be smaller than estimated above.\\

Based on these considerations we can expect the following expression
for the disruption times of star clusters due to tidal interactions in 
different extragalactic environments

\begin{equation}
t_{\rm dis}~=~  C_{\rm env} \>\left(\frac{M_{\rm
      cl}}{10^4\>\msun}\right)^{0.62}\>
  \left( \frac{\rho_{\rm amb}}{\Msun~ \rm pc ^{-3}} \right) ^{-1/2}~~
\label{eq:tdisexpected}
\end{equation}
where $C_{\rm env} \simeq 300 - 800$ Myr. 
If encounters with giant molecular clouds or disk shocking (for clusters
in orbits tilted with respect to the galactic plane) becomes important,
the disruption times might be shorter than given by 
Eq. (\ref{eq:tdisexpected}).


\section{The empirically determined  disruption times}

The disruption times of systems of star clusters in different
extragalactic environments have been determined statistically
by BL03 from samples of star clusters in four galaxies. 
We will summarize their method and results.

Suppose that the disruption times of clusters in some region of a galaxy
can be written as

\begin{equation}
\tdis = t_4 \times (\Mcl/ 10^4)^{\gamma}
\label{eq:tdisemp}
\end{equation}
where \Mcl\ is the initial cluster mass (in \Msun) and $t_4$ is the
disruption time (in yrs) of a cluster with an initial mass of 
$\Mcl = 10^4$ \Msun. 
Assume
also that the cluster formation rate is constant over the period
of time for which $t_4$ and $\gamma$ are determined, and that all clusters
had the same initial stellar IMF. 
Then, the values of
$\gamma$ and $t_4$ can be derived from the mass and age histograms of
cluster samples with a well determined brightness limit. Boutloukos \&
Lamers (BL03)
have applied this method for an empirical determination of $\gamma$
and $t_4$ in specific regions of M51, M33, (observed with $HST-WFPC2)$
and in the SMC and the solar neighbourhood.  The method has been
improved by Lamers (2004) and by Gieles et al. (2004c),
who showed that the distribution of the observed clusters in a mass
versus-age-diagram provides a more powerful method for deriving
$t_4$ and $\gamma$ than the mass and age histograms separately.

The results are given in
Table 1.  Column 2 gives the range of galactocentric distances of the
cluster sample.  Columns 3, 4 and 5 give the number of clusters and
their age and mass range. Columns 6 and 7 give the resulting values of
the disruption time $t_4$ and $\gamma$. Column
8 gives the ambient density of the host galaxy at the location of the
cluster samples used for the determination of the disruption time (see
Sect. 4).

In the determination of
$\gamma$ we adopted a cluster IMF with $\alpha=2.0$ 
(Zhang \& Fall 1999, Bik et al. 2002, de Grijs et al. 2003).  
The values of $t_4$ and $\gamma$ of the SMC is from BL03.
 The values for M51 have been redetermined by Gieles et al. (2004c) 
on the basis of the extended cluster sample of Bastian et al. (2004) 
and the improved  method mentioned above. 
The values for M33 have been redetermined by Lamers et al. (2004b), on
the basis of the extended cluster sample by Chandar et al. (2001,
2002). The new value of $t_4$ for this galaxy is considerably higher
than originally derived by BL03, due to the largely improved cluster 
sample and a redetermination of the cluster ages.
The values for the Solar neighbourhood have been redetermined by
Lamers et al. (2004b), based on the clusters sample by Loktin et al. (1994).

  Notice that the values of $\gamma$ of the four galaxies are
the same, within the uncertainty, but the values of $t_4$ differ
strongly.  The mean value of the empirically determined value of 
$\gamma$ and its uncertainty is

\begin{equation}
\langle \gamma_{\rm emp} \rangle ~=~ 0.60 \pm 0.02.
\label{eq:gammaemp}
\end{equation}
where 0.02 is the uncertainty in the mean value, which is much smaller
than the uncertainty in the individual measurements.
This exponent agrees very well with $\gamma_{\rm pred} =0.62$  
predicted by BM03,
Eq. \ref{eq:tdisBM}, on the basis of their numerical simulations. 
The agreement between the observed and the predicted
value of $\gamma$ was first pointed out by Gieles et al. (2004d), but based on
slightly different theoretical arguments.

\begin{table*}
\caption[]{ The parameters of the disruption time:
~~$\tdis = t_4 \times (\Mcl/ 10^4)^{\gamma}$}
\begin{tabular}{lcccccccl}
\hline \\
Galaxy & $r_{\rm Gal}$    & Nr        & Age range & Mass range & $\log t_4$&
$\gamma$ & log \rhoamb & Remark\\
       & kpc              & clusters  & log (yrs) & log \Msun  & log (yrs)$^a$ &
         & \Msun~pc$^{-3}$ & \\
\hline \\
 M51       & 1.0 - 5.0 &  1152 & 6.3 -- ~9.0 & 2.6 -- 5.6 &
$7.85 \pm 0.22$~(7.6)  & $0.57 \pm 0.10$ & $-0.70 \pm 0.30$ & 2\\
 M33       & 0.8 - 5.0 &  147   & 6.5 -- 10.0 & 3.6 -- 5.6 &
$8.80 \pm 0.20$~(8.1)  & $0.60 \pm 0.15$ & $-0.66 \pm 0.30$ & 3\\
 Galaxy ($d\le1$kpc) & 7.5 - 9.5 & 184 & 7.2 -- 9.5 &  --- &
$8.75 \pm 0.20$~(9.0)    & $0.60 \pm 0.12$ & $-1.00 \pm 0.04$ & 1,4\\
 SMC       & 0 - 4     &  314 & 7.6 -- 10.0 & --- &
$9.90 \pm 0.20$~(9.9)   & $0.61 \pm 0.08$ & $-2.10 \pm 0.30$ & 1\\
 & & & & & Mean &   $0.60 \pm 0.02$ & & 5\\
\hline\\
\end{tabular}

(a) The value in parenthesis is the original value from BL03 \\
(1) $t_4$ and $\gamma$ from age distribution only \\
(2) Improved determination of $t_4$ and $\gamma$ by Gieles et
al. (2004c), based on new cluster sample of Bastian et al. (2004)\\
(3) Improved determination of $t_4$ and $\gamma$ by Lamers et
al. (2004b) based on extended
cluster samples by Chandar et al.(2001, 2002)\\
(4) Improved determination by Lamers et al. (2004a) based on cluster sample of
Loktin et al. (1994)\\
(5) The quoted error is the uncertainty in the mean value

\label{tbl:1}
\end{table*}


\section{The dependence of the disruption time on the conditions
in the host galaxy}

Table 1 shows that the constant $t_4$, which is the disruption time of
a clusters of $\Mcl = 10^4 \Msun$, differs greatly between the regions
in the galaxies that were studied. This indicates that the disruption
time depends strongly on the local conditions in the host galaxy. 
In Sect. (2) we have shown that $t_4$ is expected to scale with the
inverse square-root of the mean density in the host galaxy, 
if evolutionary mass loss, relaxation driven evaporation and tidal stripping
are the dominant mass loss mechanism for clusters.
In this section we compare this predicted dependence with the
empirically derived values of the disruption time $t_4$.


\subsection{The ambient density in the host galaxies}

We estimate the ambient density \rhoamb\ at the galactocentric distances of
the observed environments for which
BL03, Gieles et al. (2004c) and Lamers et al. (2004b)
determined the value of $t_4$. 

{\bf M51}:\\
 The clusters are at galactocentric distances of 1 to
5 kpc. The column density of M51 decreases exponentially as 
$\sigma=540 \exp(-r/4.65)$ \Msunpc, with $r$ in kpc 
(Salo \& Laurakainen 2000). The total disk mass
within a ring of 1 to 5 kpc is $2.0 \times 10^{10}$ \Msun\
and the mean surface density in that annulus is $2.6 \times 10^2$ \Msunpc. 
Van der Kruit (2002) has shown that spiral galaxies have a vertical
scale-height of 1/7 times the radial scale-length. This implies
a value of $h_z=0.66$ kpc and a mean mid-plane density 
between 1 and 5 kpc of 0.20 \Msun~pc$^{-3}$. The structural
parameters of M51 derived by Athanassoula et al. (1987) result in a
very similar estimate of the mid-plane density.

{\bf M33}: \\
The sample of clusters is located at  galactocentric
distances between 1 and 5 kpc. More than 90 percent of the clusters
are between 1 and 4 kpc.
The column density of M33 decreases exponentially as 
$\sigma=525 \exp(-r/1.45)$ \Msunpc, with $r$ in kpc 
(Athanassoula et al. 1987). The total disk mass
within a ring of 1 to 4 kpc is $4.2 \times 10^9$ \Msun\
and the mean surface density is 90 \Msunpc. 
Adopting an effective thickness of the disk of $1/7$ of the radial
scale-length, i.e. $0.21$ kpc,  
we find a mean mid-plane density of $0.22 $ \Msun~pc$^{-3}$.

{\bf Solar Neighborhood}:\\ 
The mean gas and stellar density in the solar neighborhood, derived from
Hipparcos data is $0.10 \pm 0.01$ \Msun~pc$^{-3}$ (Holmberg \& Flynn 2000).

{\bf SMC}: \\
The sample of clusters is distributed throughout the SMC.
With a total mass of about $2 \times 10^9$ \Msun\ 
(Mathewson \& Ford 1984) and a radius of the central core of 4 kpc 
(Caldwell \& Laney 1991) 
we find a mean density of $8 \times 10^{-3}$ \Msun~pc$^{-3}$.

\begin{figure}
\centerline{\psfig{figure=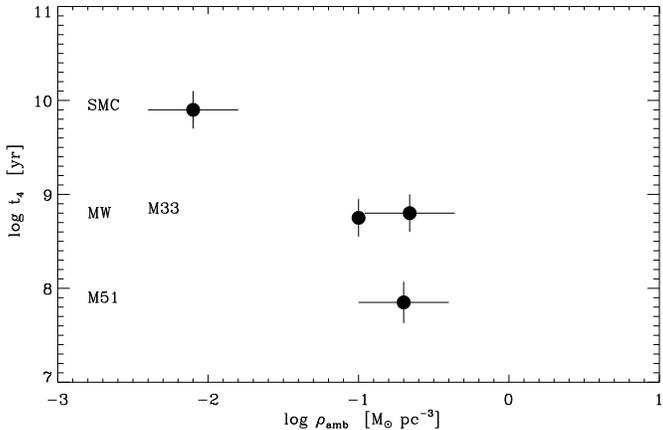,width=10.0cm}}
\caption[]{The relation between the measured values of the
 disruption time $t_4$ of clusters 
with an initial mass of $10^4$ \Msun\ and the mean ambient density  \rhoamb\
in \Msun~pc$^{-3}$, of the
galaxies at the galactocentric distances where the studied clusters were
located. }
\label{fig:1}
\end{figure}

The ambient densities are 
listed in Table \ref{tbl:1}, where we have adopted an
uncertainty of a factor 2 in the densities, except for the solar
neighbourhood. 
The relation between $t_4$ and the ambient  densities is shown in
Figure \ref{fig:1}. The figure shows a clear trend
of decreasing disruption time with increasing ambient density.

\subsection{Comparison between observed and predicted disruption time scales}

We compare the observed values of $t_4$ for the different galaxies
with the predictions based on $N$-body simulations, discussed in
Sect.2.  
These simulations
were done for clusters in Galactic environments. However, by using 
the expected scaling laws (Sect. 2),
we can apply the results of these
simulations to the conditions in the galaxies where the cluster
disruption times were measured.

\begin{figure}
\centerline{\psfig{figure=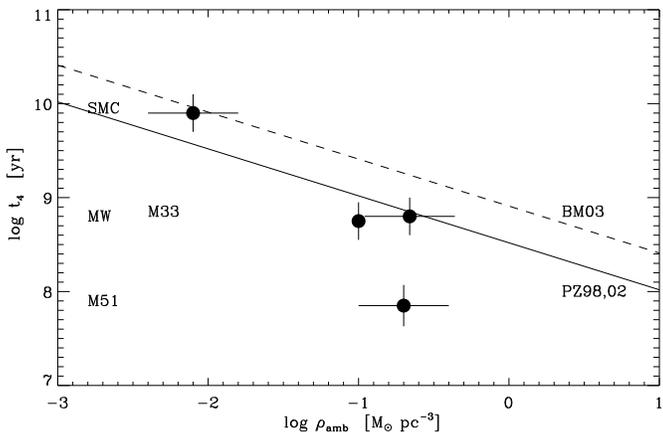,width=10.0cm,angle=0}}
\caption[]{Comparison between the observed (large dots) and predicted 
(lines) disruption time $t_4$ of clusters with an initial mass
of $10^4$ \Msun\ as a function of the mean density \rhoamb\, in
\Msun~pc$^{-3}$ of the host galaxy.
The solid curve present the predictions by Portegies Zwart et
al. (1998, 2002) and the dashed line by BM03 .
}
\label{fig:compare}
\end{figure}

In Fig. \ref{fig:compare} we compare the observed relation between $t_4$
and $\rhoamb$ with the predicted mean relations derived 
 from $N$-body simulations of
clusters of $10^4$ \Msun\ from Portegies Zwart (1998, 2002) and
BM03, shown in Fig. \ref{fig:t4predicted}.

The empirically derived values of $t_4$ roughly follow the relations
predicted by the $N$-body simulations within the uncertainty of the
data, except for the clusters in M51, whose disruption time is almost an
order of magnitude shorter than expected.

There are several possible explanations for the short cluster
lifetimes in M51, some of which originate from variations in
the birth conditions of the clusters or from star formation histories
of the host galaxy.  For example, the constant used in
Eq.\,(\ref{eq:tdisexpected}) depends via \mmean, quite sensitively on the
adopted initial mass function.  If we adopt a Salpeter
(1955)\nocite{salpeter} initial mass function with a lower mass limit
of 1\,\msun, instead of 0.1\,\msun, the mean stellar mass increases by
about an order of magnitude. Such dramatic change to the initial mass
function results via Eqs.\,(\ref{eq:tdis2}) 
 in a reduction of the
cluster lifetime by about a factor 3.  Such a lower mass cut-off for
the initial mass function is controversial, but is suggested in
young star clusters like MGG-11 (McCrady et al. 2003) and MGG-F, both
in M82 (Smith and Gallagher 2001)\nocite{Smith and gallagher} and in
several of the young star clusters in the Antennae
galaxies (Mengel et al. 2002).
 
Since M51 experienced a recent star
formation event, about 60 Myr ago 
(Bastian et al. 2004), possibly part of the discrepancy can be explained by
variations in the IMF. (This starburst event was taken into account
in the determination of the disruption time for M51 by Gieles et
al. 2004b). 
Young star formation environments also tend to
have higher metalicity, which drives stronger winds and therefore more
stellar mass loss. 
Another possible explanation for the short disruption time of clusters
in M51 is related to the strength of the tidal
field variations. M51 has the highest density contrast between the
arm and interarm regions of any spiral galaxy (Rix \& Rieke 1993), 
so the tidal field variations experienced by star
clusters will be higher than normal, which might explain the short
disruption time of the star clusters in M51.
This last effect is studied by Gieles et al. (2004a).

\section{Conclusions}

We analyzed the disruption time of star clusters in the Solar Neighbourhood, 
the SMC and in selected regions of M33 and M51. We find the following result.\\
(a) Within each region, there is a clear relation between the empirically derived 
cluster disruption time and the cluster mass of the type $\tdis
\propto M^{\gamma}$. The mean empirical value of $\gamma_{\rm emp}=0.60 \pm
0.02$ agrees very well with the predicted value of $\gamma_{\rm  pred}=0.62$, 
derived from direct $N$-body simulations of galactic star clusters.\\
(b) Comparing the disruption times in different galaxies, we find that
the disruption time depends on the ambient density in the environment
where the clusters reside. $N$-body simulations of clusters at different 
galactocentric distances predict that the disruption time decreases
with the ambient density as $\tdis \propto \rhoamb^{-0.5}$. The
empirical values of three galaxies, SMC, M33 and the Solar
neighbourhood agree roughly with this prediction. 
However, the  disruption time of star clusters in the inner 4 kpc
of M51 is shorter than predicted by about an order of magnitude.
The difference is most likely due to large tidal field variations,
due to the high density contrast in the spiral arms and interarm
regions of M51, or due the starburst triggered by the last
encounter with its companion NGC 5195. 

The disruption time of star clusters of initial mass $M_{\rm cl}$
in circular orbits in an environment with a 
mean ambient density  \rhoamb\ is given by Eq. (\ref{eq:tdisexpected}).
%
%
The disruption time may be shorter in starburst regions.


\begin{acknowledgements}

We thank Nate Bastian for discussions and comments on the manuscript,
Albert Bosma and Piet van der Kruit for advice on the
densities of spiral galaxies and Holger Baumgardt for providing the 
results of his numerical simulations.
We are grateful to the referee, Douglas Heggie, 
for constructive suggestions that improved the paper.
This work was supported by a grant from the Netherlands Research
School for Astronomy (NOVA) to HJGLML and by a grant from the 
Royal Netherlands Academy of Sciences (KNAW) to SPZ. 
\end{acknowledgements} 



\end{document}